\documentclass{PoS}
\usepackage{graphicx}

\title{Entropy Production at High Energy and $\mu_B$}

\ShortTitle{Entropy Production at High $\sqrt{s}$ and $\mu_B$}

\author{\speaker{Peter Steinberg}\\%
Brookhaven National Laboratory\\
Upton, NY 11973, USA\\
        E-mail: \email{peter.steinberg@bnl.gov}}

\abstract{
The systematics of bulk entropy production in experimental data 
on A+A, $p+p$ and $e^+ e^-$ interactions at
high energies and large $\mu_B$ is discussed.  It is proposed that
scenarios with very early thermalization, such as Landau's 
hydrodynamical model, capture several essential features of the
experimental results.  It is also pointed out that the dynamics
of systems which reach the hydrodynamic regime give similar
multiplicities and angular distributions as those calculated
in weak-coupling approximations (e.g. pQCD) over a wide range
of beam energies.  Finally, it is shown that the dynamics of
baryon stopping are relevant to the physics of total entropy
production, explaining why A+A and $e^+ e^-$ multiplicities
are different at low beam energies.  
}

\FullConference{``Critical Point and the Onset of Deconfinement'' Workshop\\
                 July 3-6 2006, Florence, Italy}

\begin{document}

\section{Introduction}

\begin{figure}[t]
\begin{center}
\includegraphics[width=75mm,angle=0]{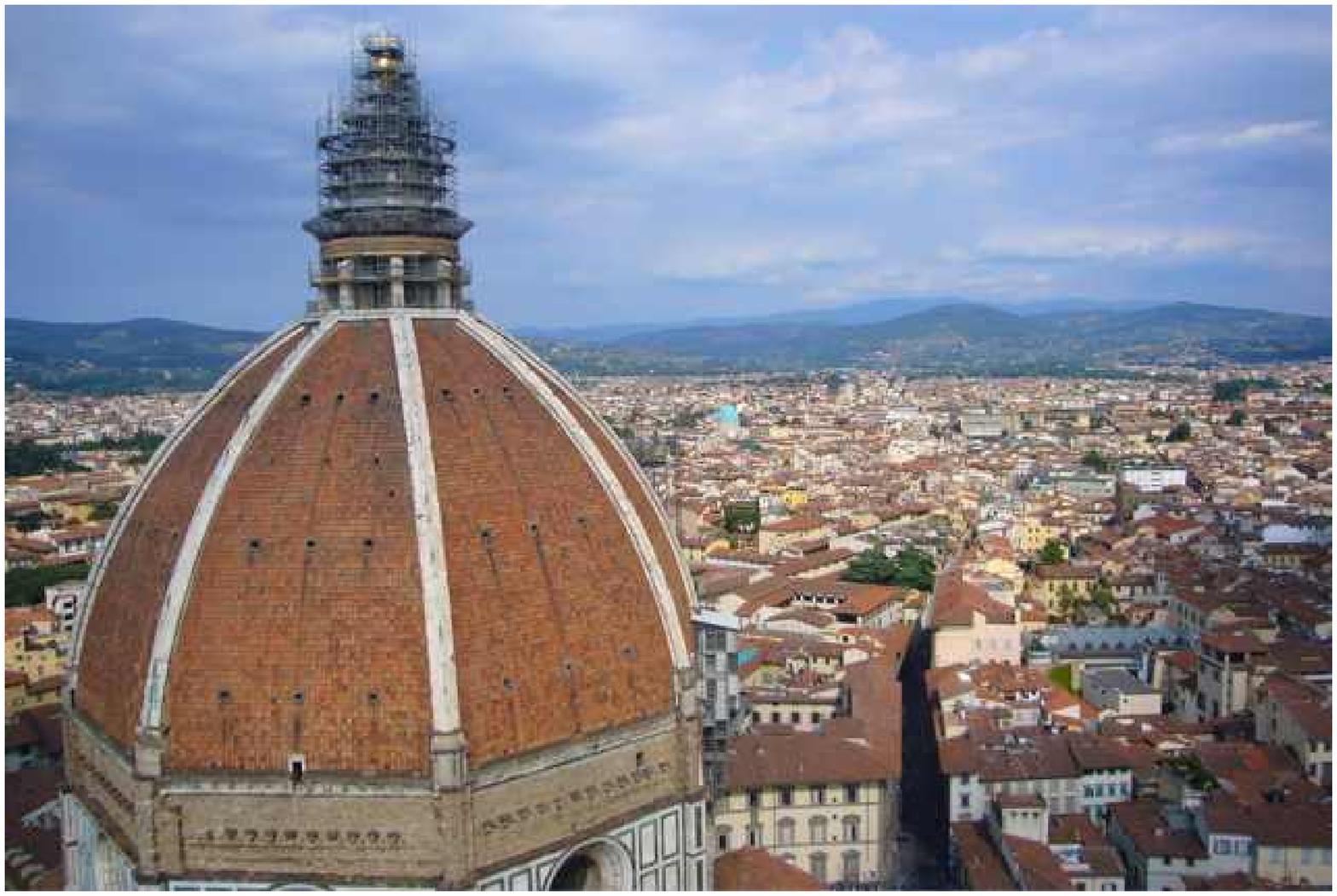}
\includegraphics[width=75mm,angle=0]{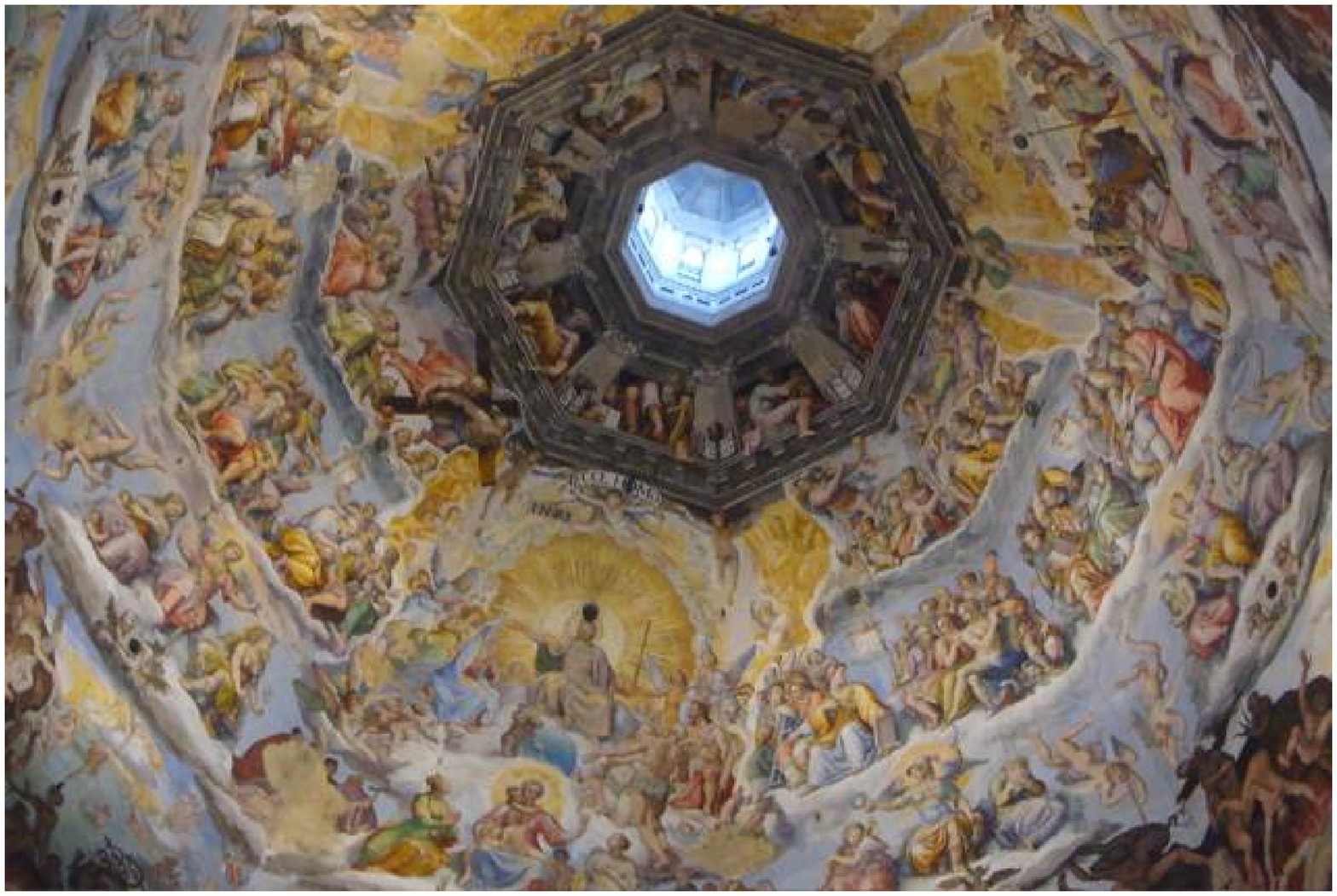}
\caption{
Brunelleschi's Duomo in Florence, seen from the outside (left) and
inside (right).  Externally, it is a simple, elegant structure,
while complex internal ``dynamics'' are seen on the inside.
\label{fig:duomo}
}
\end{center}
\end{figure}

Brunelleschi's dome was built in Florence in the 15th century and continues
to impress due to its sheer size and technical achievements.  And yet, 
examining it from the outside, you would never guess at the complex
internal ``dynamics''
of Vasari and Zuccari's interior frescoes, depicting choirs of angels,
saints, and various sins.  
The point of this work is to argue something similar about our
understanding of heavy ion collisions.  Graphical displays of RHIC events
look quite complicated, with thousands of tracks emanating from the
primary vertex composed of a variety of particles.  However, angular
distributions of inclusive charged particles can be found to show
simple ``scaling'' features, which are shared even with elementary
collisions.  Of course, just like the exterior structure provides
the support for the frescoes inside, 
these scaling patterns may well hint at the true
microscopic dynamics at work.  In any case, scaling relations 
may well suggest an overall framework into which predictions 
relevant to the upcoming Large Hadron Collider (LHC) at CERN 
and FAIR at GSI should fit.

\section{Elementary Collisions}

Compared to RHIC collisions, ``elementary'' collisions of protons and antiprotons, or electrons and positrons, which produce many hadrons seem to
look quite sparse.  Thus, they are generally thought to result
from very different physics processes.

Electron-positron annihilation into hadrons used to be
understood using concepts based on the original ``string models'' 
of the 1970's, with extensions incorporating multiple gluon 
production\cite{Sjostrand:1995iq}.
More recently, purely perturbative calculations involving gluon ladders
can capture many features of the data, down to details of
jet fragmentation.  This is especially true of calculations of the total
multiplicity, which have good descriptive and predictive power, starting
from the early SPEAR data up to the top LEP2 energies.  A full accounting
of the running coupling in jet fragmentation gives formulae that scale as
$n_{ch} \propto \alpha^A_s \exp (B\sqrt{\ln(s)})$\cite{Mueller:1983js}. 
One achieves
similar results in various ``parton cascade'' approaches, such as
JETSET, which augments the older string models with perturbative
gluon emission.  

Collisions of protons and antiprotons are generally understood in a
``two-component'' scenario.  The soft component is thought to be the
domain of ``non perturbative QCD'' but understood phenomenologically
by means of descriptive features like longitudinal phase-space and
limited $p_T$.  Various implementations of this can be tuned to
describe the available data.  Of course, jet phenomena have been 
observed as the energies increased, suggesting that there is a 
separate ``hard'' component at work in p+p collisions.  This
has been successfully modeled by combining the structure functions
measured at high-$Q^2$ in $e+p$ collisions, with pQCD cross
sections to get the angular distributions, and fragmentation 
functions measured in $e^+ e^-$ reactions used to parametrize 
the relationships between the outgoing quarks and gluons and the
measured hadrons.   

\begin{figure}[t]
\begin{center}
\includegraphics[width=80mm]{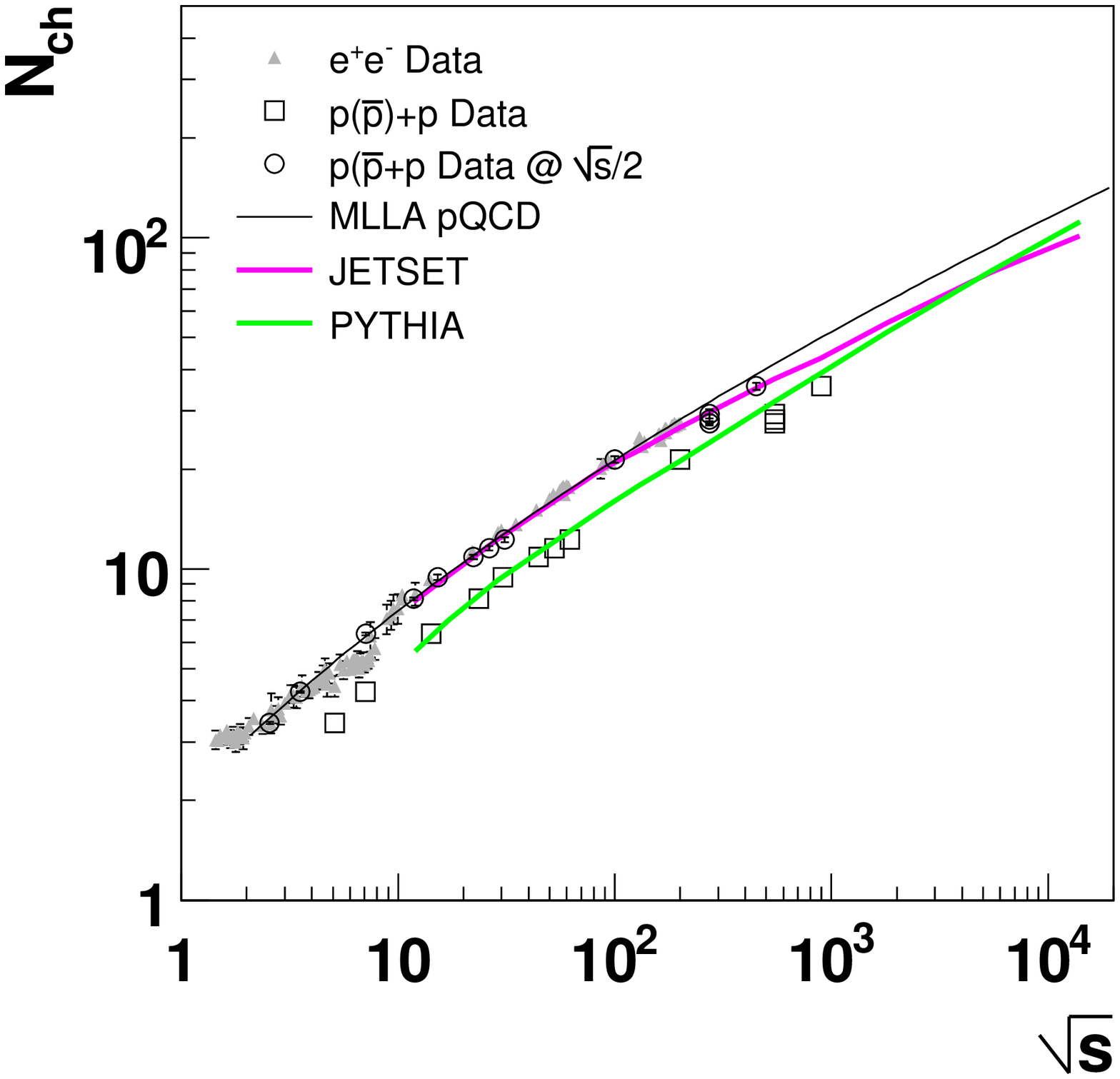}
\includegraphics[width=70mm]{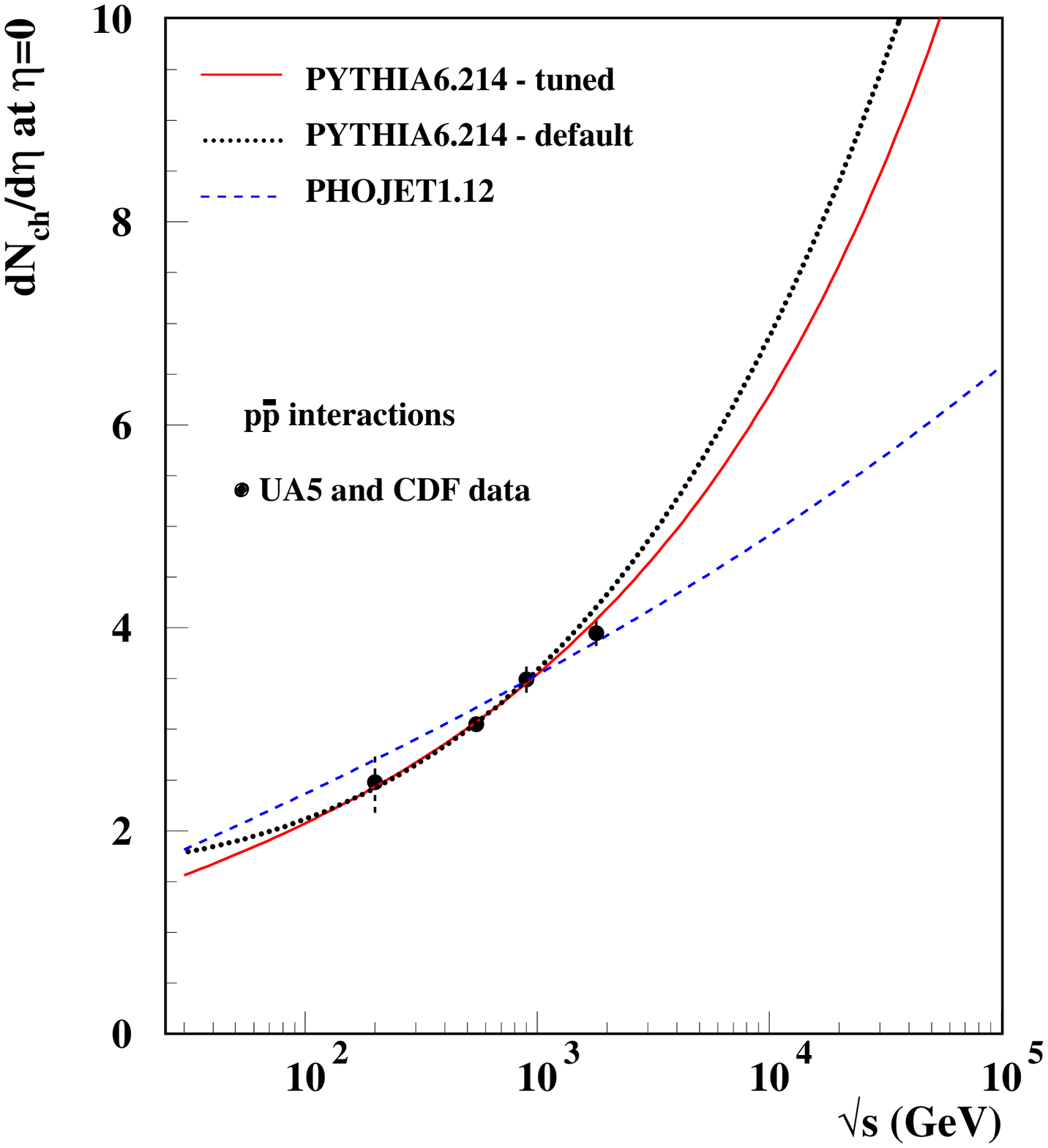}
\caption{
(left) Total primary charged particle multiplicity for $e^{+} e^{-}$ (with up to 10\% admixture of weak decays) and 
$p(\overline{p})+p$ collisions, compared with MLLA pQCD calculations, JETSET, and PYTHIA.  (right) Midrapidity density for $\overline{p}+p$ collisions vs. $\sqrt{s}$ compared with PYTHIA and PHOJET, from Ref. \cite{Moraes:2006}
\label{fig:emub_total_4a}
}
\end{center}
\end{figure}

Various models incorporate the hard and soft 
components in different schemes, such as PYTHIA\cite{Sjostrand:2006za}, 
HERWIG\cite{Marchesini:1991ch}, PHOJET~\cite{Engel:1994vs}, and HIJING\cite{Gyulassy:1994ew}.  And yet, despite being based on similar
inputs, most of these models predict different extrapolations
of existing data to high energies~\cite{Moraes:2006}, 
as shown in Fig.~\ref{fig:emub_total_4a}.  
One expects an interesting
early running of the LHC while the various models (or tunings
thereof) are validated, or ruled out, by the first data.

\begin{figure}[t]
\begin{center}
\includegraphics[width=70mm]{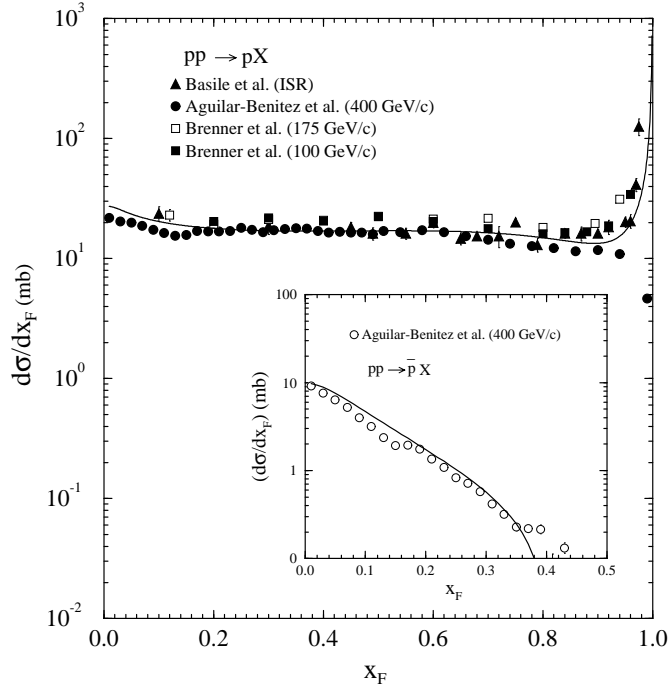}
\vspace*{5mm}
\caption{
Inclusive $dN/dx_F$ for protons $p+p$ collisions at several energies.  Inset shows the same for anti-protons.  From Ref.~\cite{Batista:1998ry}.
\label{fig:figure1}
}
\end{center}
\end{figure}

One major uncertainty in understanding soft particle production
in p+p (and $\overline{p}+p$) collisions is related to the lack
of dynamical mechanisms in the models.  It is still not generally
understood how the incoming baryons are ``stopped'', and their 
energy transmuted into particles\cite{Busza:1983rj}.  
The net rapidity loss of the incoming baryons has been
studied extensively in fixed-target experiments as well as at the
ISR (but not at the Tevatron collider, unfortunately)~\cite{Basile:1980ap,Brenner:1981kf}.  It has been found
that the distribution of $x_F = 2p_Z/\sqrt{s}$, the fraction of
energy found in the outgoing
``leading'' particles is essentially flat (but with a quasi-elastic
peak near $x_F \sim 1$)\cite{Batista:1998ry}, as shown in Fig.~\ref{fig:figure1}.
More interestingly, this net baryon rapidity
loss is found to correlate strongly with the total multiplicity,
and approach the $e^+ e^-$ multiplicity measured at the same 
$\sqrt{s}$\cite{Basile:1980ap}. 
In fact, the $e^+ e^-$ and $p+p$ data overlap each other if
$p+p$ is plotted at $\sqrt{s_{eff}}=\sqrt{s}/2$.
This suggests that 1) the net baryons measured in p+p
collisions reflect the inelasticity of the collision,
and 2) the basic mechanisms of total entropy production in
both $e^+ e^-$ and $p+p$ are quite similar.

One way to understand the similarity between the entropy produced
in these two systems is by simply postulating that both are
the result of an equilibration process.  Cooper, Frye and 
collaborators~\cite{Cooper:1974ak}
worked under this assumption in the 1970's.  Thus, whatever 
complicated dynamics might be different between the two systems
is rendered irrelevant by strong interactions between the
fundamental constituent degrees of freedom.  In that scenario,
the rest of the evolution is isentropic and simply expresses
the total entropy via the total multiplicity.
Clearly, this is a difficult scenario to 
consider if one conceives of it proceeding via the kinetic equilibrium 
of the outgoing particles.  However, it seems less problematic
if the particles are thought
to be the consequence of the freezeout of a fluid with many
strongly-interacting degrees of freedom into the thousands of available
mass states of QCD.  This is the
model that Fermi~\cite{Fermi:1950jd} and Landau~\cite{Landau:gs}
inadvertently proposed in the 1950's.

\section{Landau's Hydrodynamical Model}

Fermi and Landau both arrived at a simple formula for the total
multiplicity in the early 1950's\cite{Fermi:1950jd,Landau:gs}.  
The derivation simply assumes
complete thermalization of the total energy $E=\sqrt{s}$ in a
Lorentz-contracted volume $V=V_0/\gamma=V_0/(\sqrt{s}/2m_N)$,
leading to an initial energy density $\epsilon = s/2m_N V_0$, which
increases quadratically with $\sqrt{s}$.  Assuming the blackbody
equation of state $p=\epsilon/3$ and the first law of thermodynamics
$d\epsilon = Td\sigma$, leads to a scaling of the entropy density 
as $\sigma \propto s^{3/4}$.  
Multiplying the entropy density by the volume gives a total
entropy $S=\sigma V \propto s^{3/4}/s^{1/2} \propto s^{1/4}$.
This is the famous Landau-Fermi multiplicity scaling formula, which
suggests that total multiplicities will scale as the square root
of the CMS energy, $N \propto \sqrt{E_{cm}}$.
For a more generic equation of state (e.g. $\epsilon = c^2_s p$), 
$N \propto (1/2)(1-c^2_s)/(1+c^2_s)$~\cite{Cooper:1974ak}.

\begin{figure}[t]
\begin{center}
\includegraphics[width=70mm]{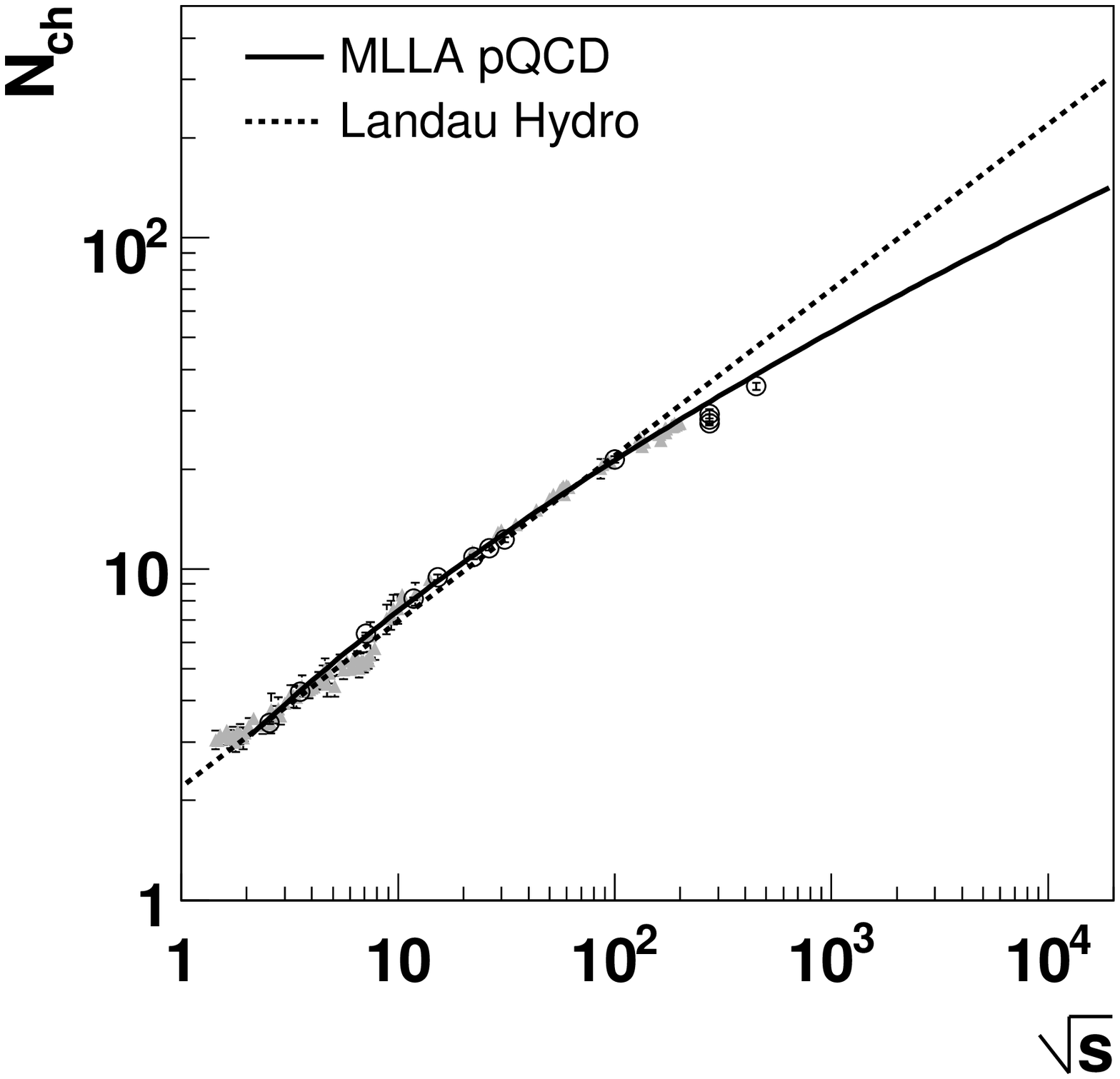}
\includegraphics[width=80mm]{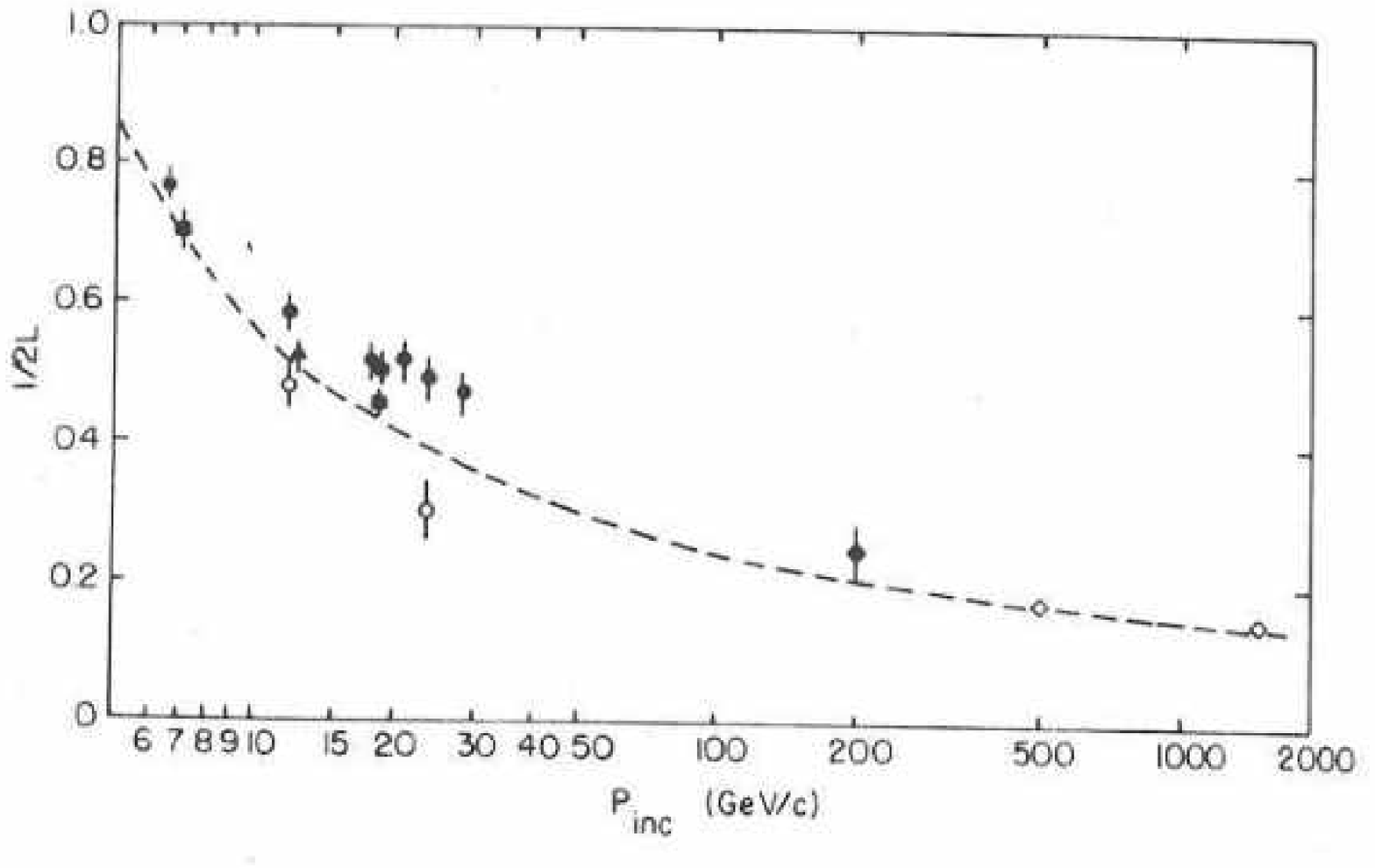}
\caption{
(left) $N_{ch}$ for $e^{+} e^{-}$ and $p(\overline{p})+p$ at $\sqrt{s_{eff}}=\sqrt{s}/2$ compared with MLLA pQCD and Landau Hydrodynamics.  Both functions have been adjusted by an overall scale factor. (right) Fits to $1/2L$ for $p+p$ data over a range of energies, from Carruthers and Duong-Van~\cite{Carruthers:1973ws}.
\label{fig:emub_total_5a}
}
\end{center}
\end{figure}

While the pQCD formula mentioned above, shown
in Fig.\ref{fig:emub_total_5a},
does a good job for the $e^+ e^-$ data, the Landau-Fermi formula
does an equally good job describing the high-energy data
when tuned on the lower energy data.  
It also naturally explains the constant ratio between the $p+p$
and $e^+ e^-$ data at the same $\sqrt{s}$, since 
$s^{1/4}_{eff}=\sqrt{1/2}\sim 70\%$.
Of course, it remains an
open question how higher-energy data will turn out, given that
the two formulae differ significantly at much higher energies
(pQCD giving $N_{ch} \sim 100$ and Landau giving $N_{ch} \sim 160$ at
LHC energies)
and the $\overline{p}+p$ data already seems to trend below
even the pQCD prediction shown above.

Of course,
the dynamical evolution does not end with the initial equilibrated
system postulated by the Landau-Fermi model.  Landau correctly
recognized that if such a system achieves local
equilibrium (i.e. with vanishingly-small mean free paths), it
will behave {\it hydrodynamically}~\cite{Landau:gs,Belenkij:cd}.
The blackbody EOS implies a 
locally-traceless stress energy tensor, and thus scale-free (i.e. conformal)
dynamics.  Thus, the evolution of the system is determined only
by the scales imposed at the beginning (the energy and volume)
and at the end (the familiar freezeout condition such that
evolution stops at $T=T_{ch}$)\footnote{This is not dissimilar to
QCD calculations, which take as input a hard scale $Q$ and a
self-generated cutoff $\Lambda_{QCD}$.  In fact, there are many
intriguing similarities between hydrodynamics and field theory
calculations, as pointed out by Carruthers\cite{Carruthers:dw}}

Landau's well-known initial conditions are quite simple:
an enormous energy density with no longitudinal motion,
packed into a volume contracted along the z axis by $1/\sqrt{s}$.  Following
the evolution analytically from its initial 1+1D expansion
to the late 3+1D expansion (using various approximations
along the way), he found that the rapidity
distribution of the fluid elements at freezeout is described by
a Gaussian distribution with 
variance $\sigma^2_y = (1/2)\ln(s/4m^2_N) \equiv L$.
Cooper, Frye and Schonberg completed the modern interpretation
of hydrodynamics by suggesting that the
fluid elements are not particles but
hadronic fireballs which decay isotropically in their own
rest frame\cite{Cooper:1974ak}.  Carruthers and Duong Van
found that Landau's model was a better fit to data than
the boost-invariant scenarios popular at the time~\cite{Carruthers:1973ws},
as shown in Fig.\ref{fig:emub_total_5a}.

It is worth taking a few moments to remark on 
what the Landau model {\it is}:  It is
a 3+1D model which assumes rapid local equilibration and has no
free parameters.  It has two scales, $\sqrt{s}$ and $T_{ch}$ which
determine the initial and final states.  Finally, it describes
the energy dependence of the produced entropy and its angular
distributions.  There is no nuclear transparency in the model and
no assumption of boost-invariance.  Rather, the entire system
is explicitly assumed to be in local thermal contact on asymptotically
small time scales $t_0 ~ 1/\sqrt{s}$ as the beam energy increases.

And a few words on what the Landau model {\it is not}:  There is
no description of net-baryon dynamics (or those of any conserved
charges).  There is no phase transition, but just a single 
EOS $p=\epsilon/3$.  There is no hadronization {\it per se}, but
just a simple freezeout criterion ($T=T_{ch}$), and thus no
mass dependence of $dN/dy$ (something which was discussed in
the 1970's by Cooper and Frye) and certainly no resonance decays.
Since these are clearly important pieces of physics, clearly
seen in data, these issues should be seen as caveats for
the various conclusions drawn later.

\begin{figure}[t]
\begin{center}
\includegraphics[width=80mm]{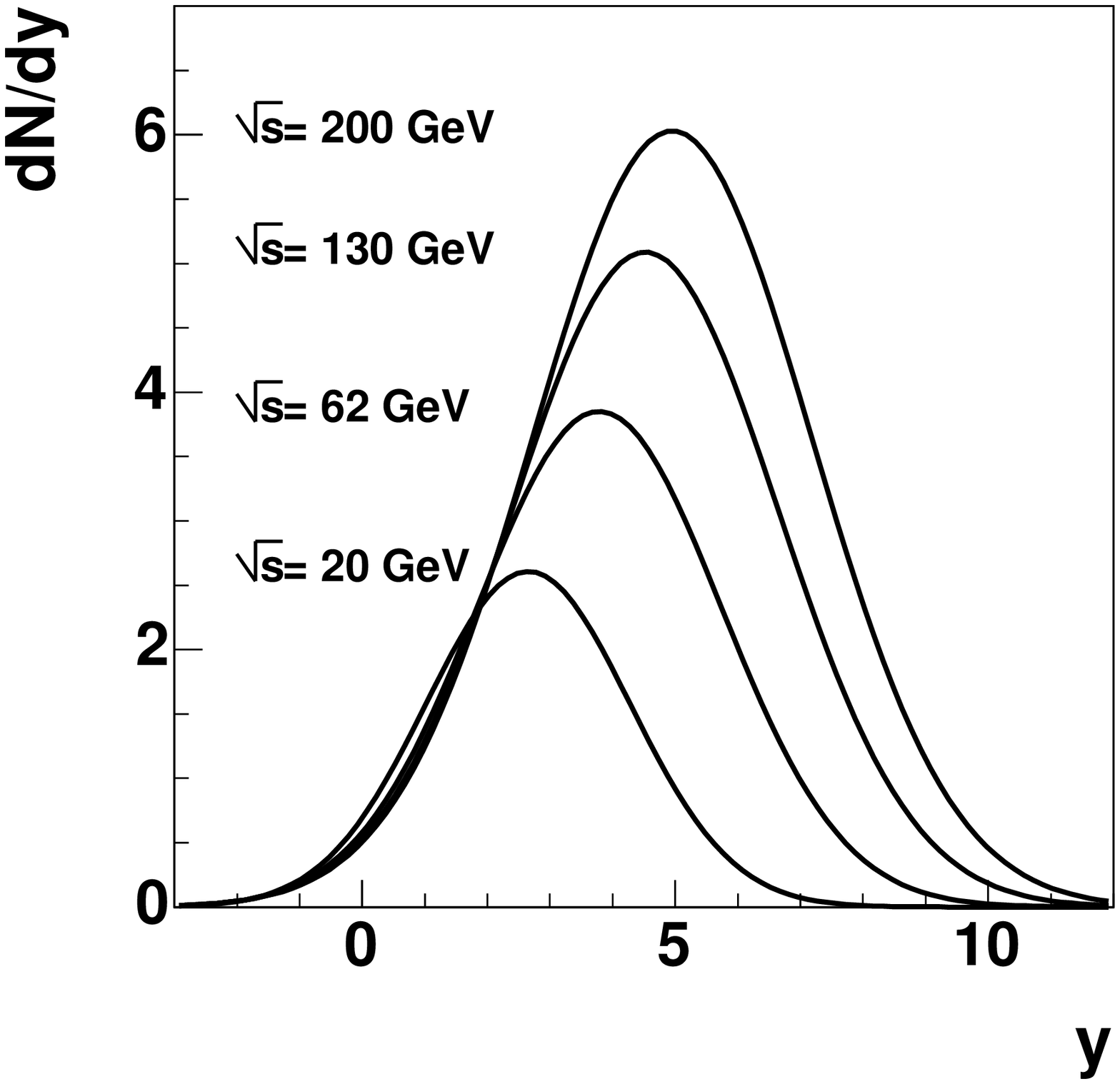}
\includegraphics[width=70mm]{WP21b_dNdeta_etaMNSybeam_pp_ee.eps}
\caption{
(left) Calculations of $dN/dy$ 
(with an arbitrary overall scale) the Landau hydrodynamic model (Equation 3.1), 
seen in the fixed target frame, showing ``extended longitudinal scaling''. (right) $dN/dy_T$, the rapidity distribution along the thrust axis in $e^{+} e^{-}$ collisions, viewed in the frame of the outgoing quark ($y_T - y_{jet}$), from Ref.\cite{Back:2004je}.
\label{fig:landau_limfrag_2}
}
\end{center}
\end{figure}

One very non-trivial feature of Landau's hydrodynamical model
appears when it is combined with the Landau-Fermi multiplicity
formula
\begin{equation}
\label{landaulimfrag}
\frac{dN}{dy}=K s^{1/4} \frac{1}{\sqrt{2\pi L}} 
\exp \left( -\frac{y^2}{2L} \right)
\end{equation}
and then viewed in the rest frame of one of the projectiles by
making the transformation $y'=y-y_{beam}=y-(L+\ln(2))$, where $L=\ln (\sqrt{s}/2m_P)$\,
\begin{equation}
\frac{dN}{dy^\prime} \sim \frac{1}{\sqrt{L}} \exp \left( -\frac{y^{\prime 2}}{2L} - y^{\prime} \right)
\end{equation}

One sees that this is approximately a function of $y^\prime$ alone, 
especially near $y^{\prime}=0$, with 
some slight scale breaking.  A direct plot of $dN/dy^{\prime}$ at
several beam energies, shown in the left panel of
Fig.~\ref{fig:landau_limfrag_2}, 
shows the phenomenon of ``limiting fragmentation''
or ``extended longitudinal scaling''~\cite{Back:2002wb} even more clearly.  
While not an original observation about the Landau model
(see e.g. Ref.\cite{Carruthers:1973ws}), this
scaling was rediscovered in this context in Ref.\cite{Steinberg:2004vy}.
This is a non trivial outcome of the formulae, and is even more intriguing
considering that it is clearly seen in both $p+p$ and $\overline{p}+p$
data with respect to the beam axis, as well as $e^+ e^-$ data with
respect to the thrust axis~\cite{Back:2004je} as shown in the right panel of
Fig.~\ref{fig:landau_limfrag_2}.

\begin{figure}[t]
\begin{center}
\includegraphics[width=140mm]{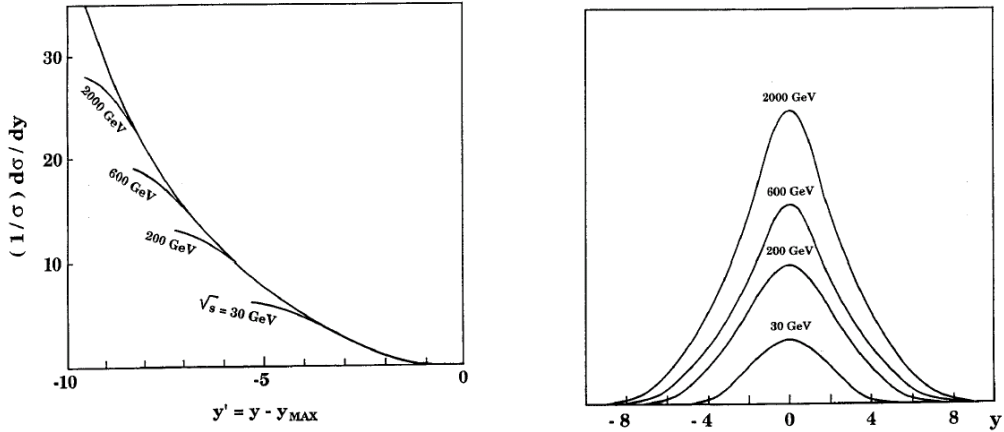}
\caption{
MLLA pQCD calculations, from Ref.\cite{Tesima:1989ca}, showing 
rapidity distributions in the rest frame of the outgoing quark (left)
and in the CM frame (right).
\label{fig:tesima}
}
\end{center}
\end{figure}

But the surprises of Landau's model are not just limited to its
relevance to experimental data.  The calculations of jet fragmentation
in perturbative QCD, in the MLLA framework discussed above, have been done by 
several authors during the 1980's.  In Ref.\cite{Tesima:1989ca},
Tesima performed MLLA pQCD calculations (which
have a different anomalous dimension than Mueller's, and thus
presumably a different energy dependence) for
the rapidity distribution of emitted gluons.  He found that the
rapidity was approximately gaussian with a width scaling as $\sqrt{\ln(s)}$
and ``translational invariance'', seen by observing the fragmentation
functions as a function of $y^{\prime}=y-y_{max}$.  Finally, we have
already seen that the MLLA formula gives similar multiplicities to the
Landau-Fermi formula over energy ranges for which data exist.  Thus,
we find that, even parametrically, pQCD and the Landau model can give similar
results.  Whether this is a particularly ornate accident, or whether
the mathematics (non-Abelian gauge theories, and 3+1D hydrodynamics with
Landau's initial conditions and freezeout criterion) share a deep
underlying structure seems to be a particularly intriguing question.  

The prevalence of extended longitudinal scaling in elementary 
collisions, and the predictions of this phenomenon from both
MLLA pQCD and Landau's hydrodynamical model should not be
forgotten when discussing the phenomenon in A+A in the context
of newer theoretical frameworks, some of which will be discussed
in the next section.  The theoretical predictions for this scaling
should also be kept in mind when trying to predict the shape
of $dN/d\eta$ and the value of $dN/d\eta(\eta=0)$, e.g. in 
Ref.\cite{Busza:2004mc}.  While the data
suggests a ``linear'' trend to the limiting curve, the models
shown here (both pQCD and Landau) suggest a nonlinearity as the
energy increases, as seen in Figs.\ref{fig:landau_limfrag_2} and
\ref{fig:tesima}.

\section{Heavy Ion Collisions}

\begin{figure}[t]
\begin{center}
\includegraphics[width=95mm]{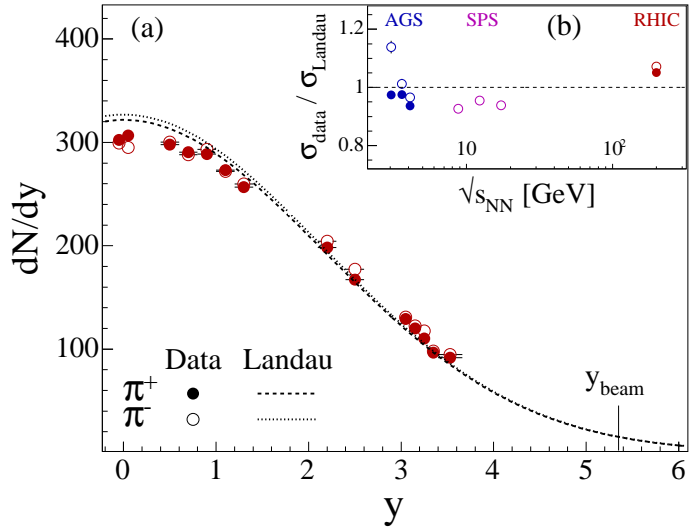}
\caption{
Rapidity distribution of pions measured in A+A by BRAHMS.
The inset shows the comparisons of Gaussian fits over a wide
range of $\sqrt{s_{NN}}$ divided by the Landau 
prediction $\sigma_{Landau} = \log(\sqrt{s}/2m_P)$.
\label{fig:landau}
}
\end{center}
\end{figure}

Moving from elementary collisions to heavy ion collisions brings in a
large number of new dynamical considerations.  The initial state 
should be characterized by shadowed parton distribution functions, as
well as the nuclear geometry suggested by Glauber calculations.  The
early dynamics are driven by hard parton scattering and subsequent
reinteractions, possibly leading to equilibration.  Eventually the
momentum transfers become low enough that hadron formation is
preferred, and the quark chemistry freezes out, incorporating the
thermalized quarks as well as the ones from jet fragmentation.  
These hadrons themselves may rescatter if the densities are sufficiently
high, leading to an eventual thermal freeze-out.  Finally, 
the final-state hadrons themselves decay, either immediately via
strong processes, or over macroscopic distances via weak processes.
All of these stages are in principle independent of the others,
and thus could lead to
a non-trivial energy and geometrical dependence as the relative
contributions of soft and hard processes change 
(e.g. HIJING~\cite{Gyulassy:1994ew}) as well as rescattering
in the partonic and/or hadronic phases \cite{Lin:2004en}.

\begin{figure}[t]
\begin{center}
\includegraphics[width=95mm]{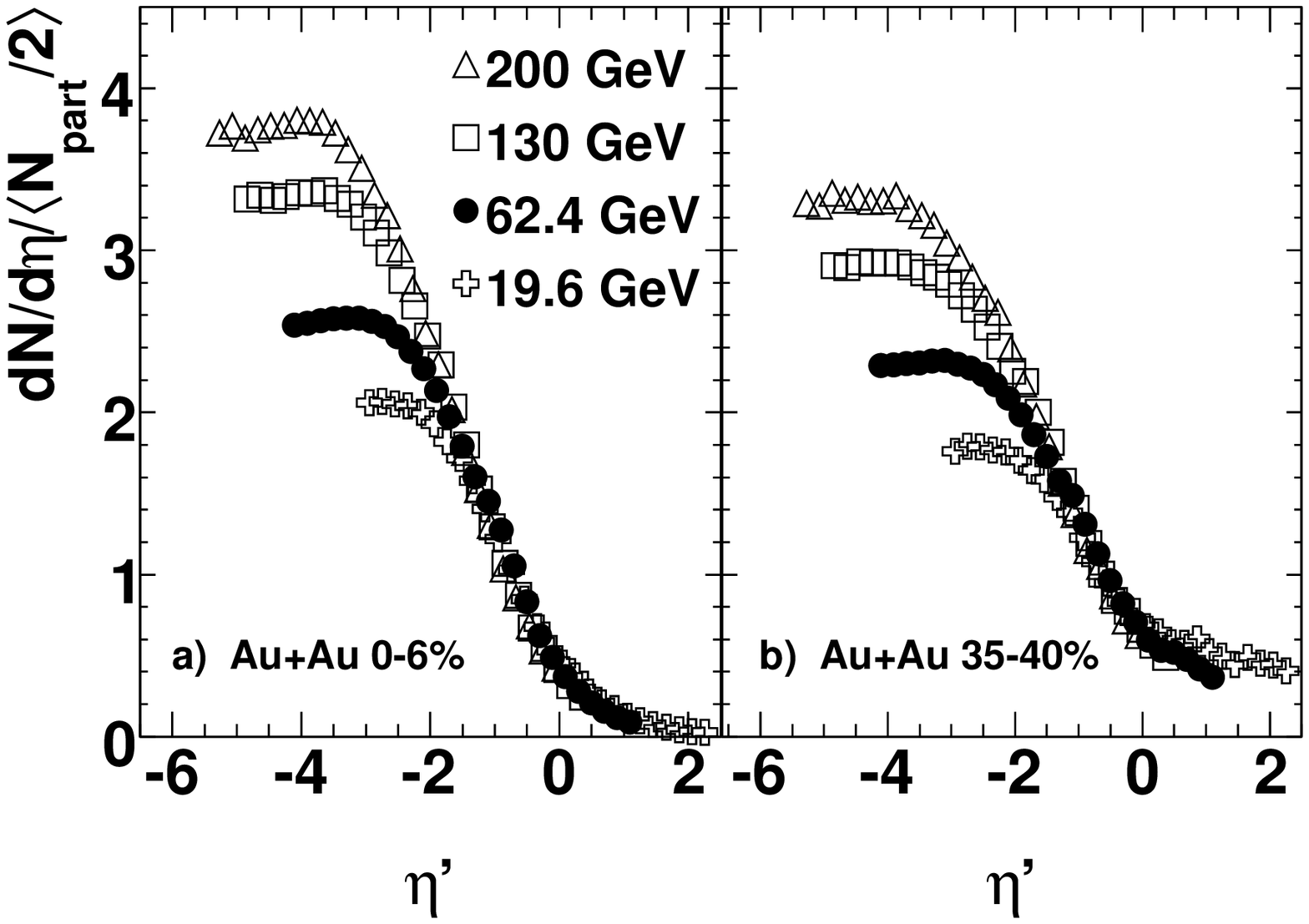}
\includegraphics[width=55mm]{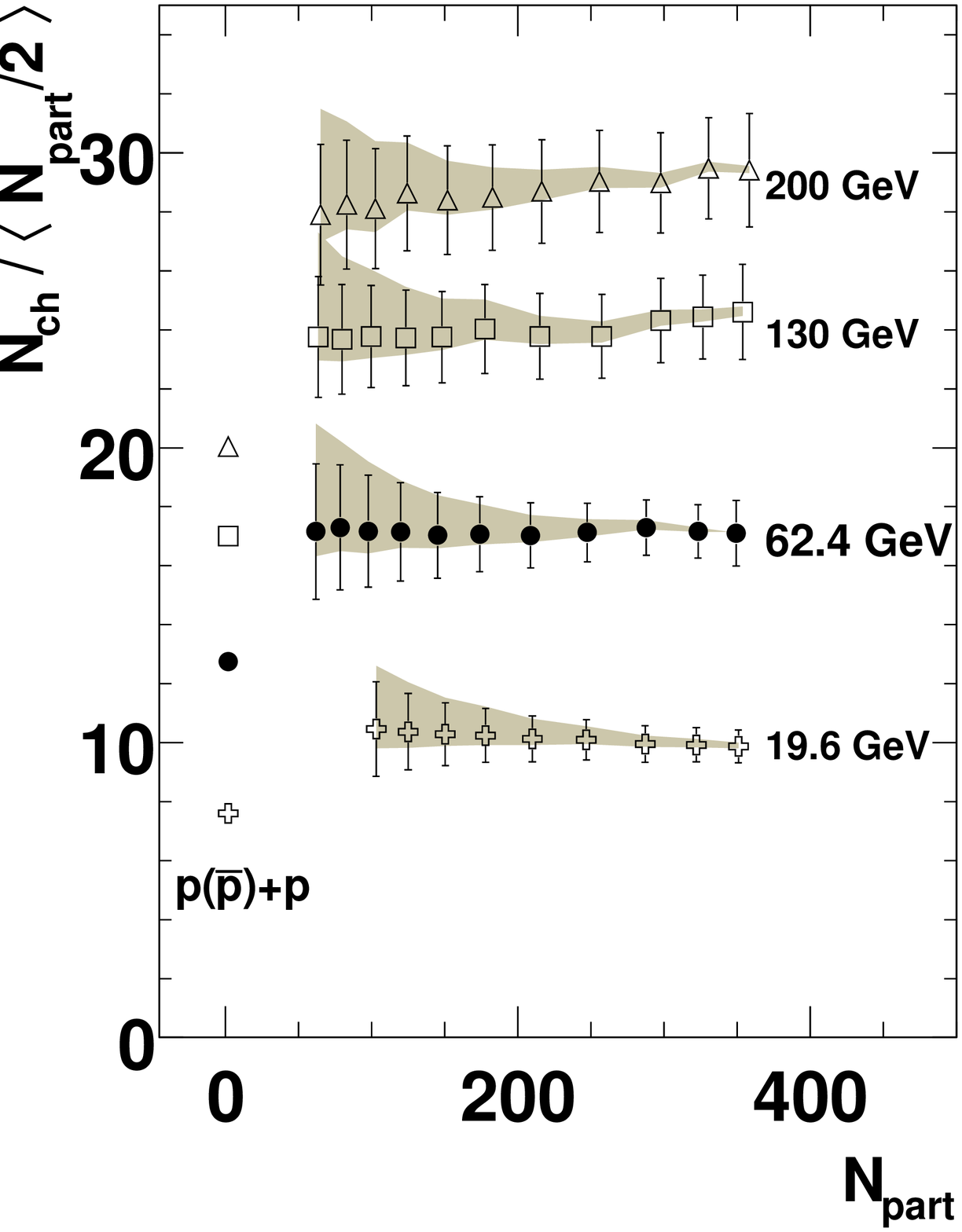}
\caption{
(left) Extended longitudinal scaling for Au+Au collisions at RHIC, from
Ref.\cite{Back:2005hs}.  (right) Total multiplicity divided by the
estimated number of participant pairs, as a function of $N_{part}$,
from Ref.\cite{Back:2005hs}.
\label{fig:mult63_lf}
}
\end{center}
\end{figure}

Surprisingly, the simple behavior in particle multiplicities seen
in $p+p$ and $e^+e^-$ collisions are quite similar to what is
actually found in A+A collisions.  
The rapidity distributions of pions measured by BRAHMS~\cite{Bearden:2004yx}
and experiments at lower energies appear to be Gaussian, as
seen in Fig.~\ref{fig:landau}, and
approximately follow the Landau predictions from 1955.
The angular distributions at RHIC also
clearly show extensive longitudinal scaling~\cite{Back:2005hs}, 
as shown in Fig.~\ref{fig:mult63_lf}.
More interestingly, the limiting curve is wider in more peripheral events,
with a long tail extending to large $\eta$ (which may be partially
explained by the presence of spectators with substantial $p_T$, but
not completely).
Combining this fact with the decreasing multiplicity near mid-rapidity
in more peripheral events, it is found that the overall multiplicity
(subtracting part of the tail using phenomenological fits to extend
into the unmeasured region) is approximately constant when
dividing out by the number of participant pairs ($N_{part}/2$).
This is shown in Fig.~\ref{fig:mult63_lf} for a wide range of energies 
measured at RHIC~\cite{Back:2005hs,Back:2006yw}.

\begin{figure}[t]
\begin{center}
\includegraphics[width=110mm]{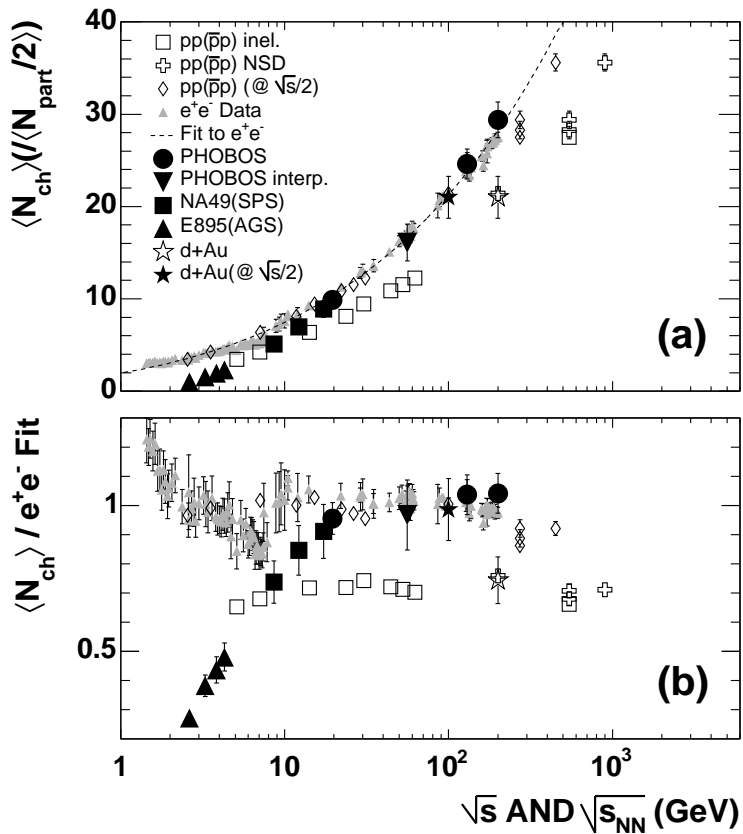}
\caption{
(top) Compilation of the total multiplicity for various systems,
from Ref.\cite{Back:2006yw}. (bottom) Data divided by the MLLA-based
fit to the $e^{+} e^{-}$ data.
\label{fig:total_ratio_inelNSD_da}
}
\end{center}
\end{figure}

The absence of a strong centrality dependence makes it possible to
compare $N_{ch}/(N_{part}/2)$ in A+A with other systems~\cite{Back:2006yw}.
As discussed above,
the $p+p$ data is similar to the $e^+ e^-$ if one takes the
$\sqrt{s}$ to be an effective $\sqrt{s_{eff}}=\sqrt{s}/2$, 
accounting for the average $x_F$ of the leading particles.
This is assuming that the flat $dN/dx_F$ distribution
is mainly comprised of baryons that do not participate in the
thermalization or subsequent dynamical evolution.  
Conversely, it is found that A+A and $e^+ e^-$ data
are similar to one another between $\sqrt{s}=20-200$ GeV 
without any other adjustments except 
dividing by $N_{part}/2$~\cite{Back:2006yw}, as shown in
Fig.\ref{fig:total_ratio_inelNSD_da}.
Given the previous comparisons of $p+p$ and $e^+ e^-$, one
particular efficient way to understand the comparisons with A+A
is to postulate that the multiple collisions experienced by
each participant (as $\nu > 2-3$ for all centralities considered
in Ref.~\cite{Back:2006yw}) essentially stops all of the incoming
energy.  This alleviates the leading particle effect, and thus
one finds the multiplicity per participant pair to be ``universal''
without additional scaling.

These comparisons are 
somewhat mysterious if one considers $e^+ e^-$ reactions as involving
just the physics of perturbative gluon radiation, while A+A 
is usually discussed in terms of 
a strongly-interacting partonic medium~\cite{Gyulassy:2004zy}.
These two appear at first glance to 
be completely opposite limits of QCD physics, the very hard
and very soft.  However, it was mentioned above that 
parametrically, MLLA pQCD and Landau hydrodynamics are quantitatively
very similar in their output, even if they do not appear to have
similar functional forms.  
The Fermi-Landau scenario would also
naturally predict that the multiplicity should scale linearly with
the initial volume, which is clearly compatible with (and essentially
predicted) the linear scaling of the total multiplicity with
$N_{part}$ shown above.
The same angular distributions as a function of $\sqrt{s}$ 
predicted by Landau also seem to appear
in the elementary collisions as well.
Perhaps it is not
necessary to use heavy nuclei to achieve local equilibration.
It should be kept in mind 
that only the first radiations in a hard process are at a truly hard scale.
Subsequent gluon emissions require the summations of more-and-more 
complicated many-gluon diagrams, which perhaps drive the
final distributions toward something resembling local equilibrium.

\section{Some Predictions for the LHC}
\begin{figure}[t]
\begin{center}
\includegraphics[width=100mm]{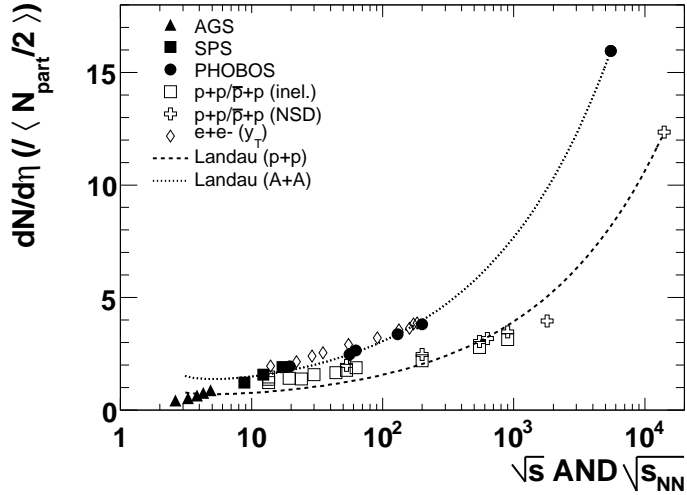}
\caption{
Midrapidity density for A+A, $p+p$ and $e^+ e^-$ as a function of
$\sqrt{s}$ and $\sqrt{s_{NN}}$.  Calculations of the Landau predictions
are superimposed, each fit to high energy A+A and $p+p$ data, with
points showing the LHC predictions made in this work.  The A+A calculations
agree reasonably while the $p+p$ fit is less satisfactory over the
full range.
\label{fig:midrap}
}
\end{center}
\end{figure}

It seems natural here to try and make predictions for the LHC
with the Landau model.
From the basic formula,
the midrapidity density scales as $\rho_0(\sqrt{s}) \propto s^{1/4}/ \sqrt{\ln
\left( \sqrt{s}/2m_{P} \right)}$ -- with no free parameters.
Thus, the ratio of $\sqrt{s}=14$ TeV to $\sqrt{s}=200$ GeV in 
proton-proton collisions will be 
$\rho_0(14 TeV)/\rho_0(200 GeV) \sim 6.1$.  
The ratio of $\sqrt{s_{NN}}=5.5$ TeV to $\sqrt{s_{NN}}=200$ GeV
for A+A (where $\rho_0$ is scaled by $N_{part}/2$)
will be $\rho_0(5.5 TeV)/\rho_0(200 GeV) \sim 4.0$.  
This is shown in Fig.\ref{fig:midrap}, which includes $\rho_0$ for
several types of collisions.  Fits of the Landau energy dependence
to data of each type (RHIC data for A+A,
NSD UA5 data for $\overline{p}+p$) have been made, to account for 
the different $p_T$ distributions as well as the overall multiplicity 
scale.

It is interesting that while the formula gets the higher energy
RHIC data (and section 5 will attempt and explanation for the
lack of agreement at lower energies), the description 
of the $p+p$ and $\overline{p}+p$ is less 
satisfactory, even qualitatively.  Unfortunately, there may be
several factors which could lead to this.  Considering yields
at mid-rapidity makes comparisons more sensitive to the details
of particle production, both species and $p_T$ dependence.
There are also issues to do with triggering, especially the
contribution from diffractive events, which are not well
understood theoretically, and are difficult to control experimentally
unless one is actively measuring leading particles.
These factors would certainly complicate a trivial application of the 
Landau formula for $dN/dy$
in a limited region of $dN/d\eta$.  Clearly, the LHC will be an
interesting place to test these ideas over a large range of $\sqrt{s}$.

\section{Total Multiplicity at Finite $\mu_B$}

And yet, it is clear that while all of the systems are similar
between $20-200$ GeV, the heavy ion data is systematically below
the $e^{+} e^{-}$ and $p+p$ data below 20 GeV, with progressively
larger deviations with decreasing beam energies.  This
might suggest that the ``universality'' of bulk particle production
described in Section 4 (and explained here by the surprising relevance of 
Landau's hydrodynamical model) is broken at lower energies.
However, a natural explanation of these data can be constructed
by considering the role of the incoming baryons, an important 
dynamical issue ignored in Landau's papers.  The following discussion
is based primarily on Ref.\cite{Cleymans:2005km}.

\begin{figure}[t]
\begin{center}
\includegraphics[width=80mm]{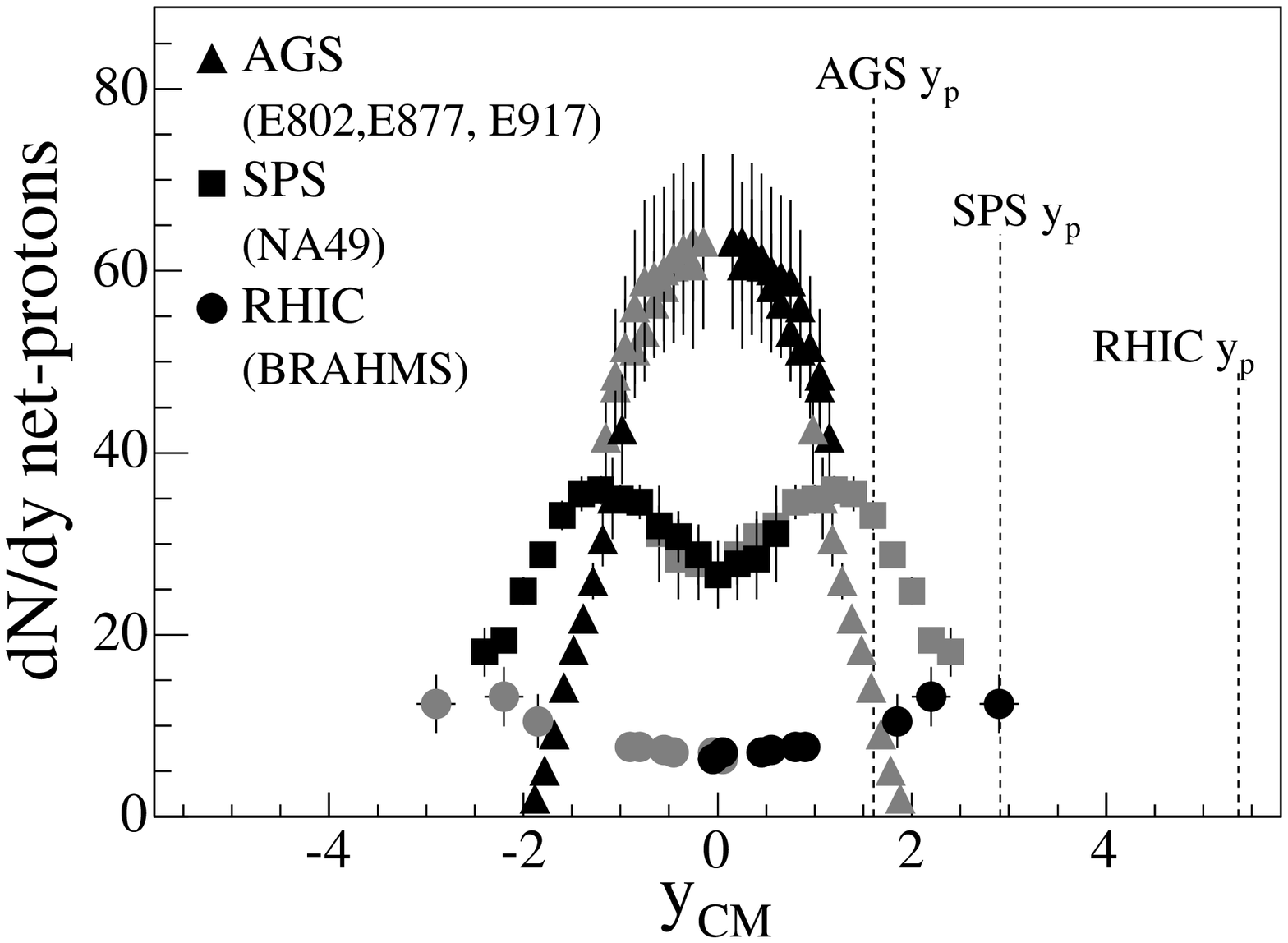}
\includegraphics[width=65mm]{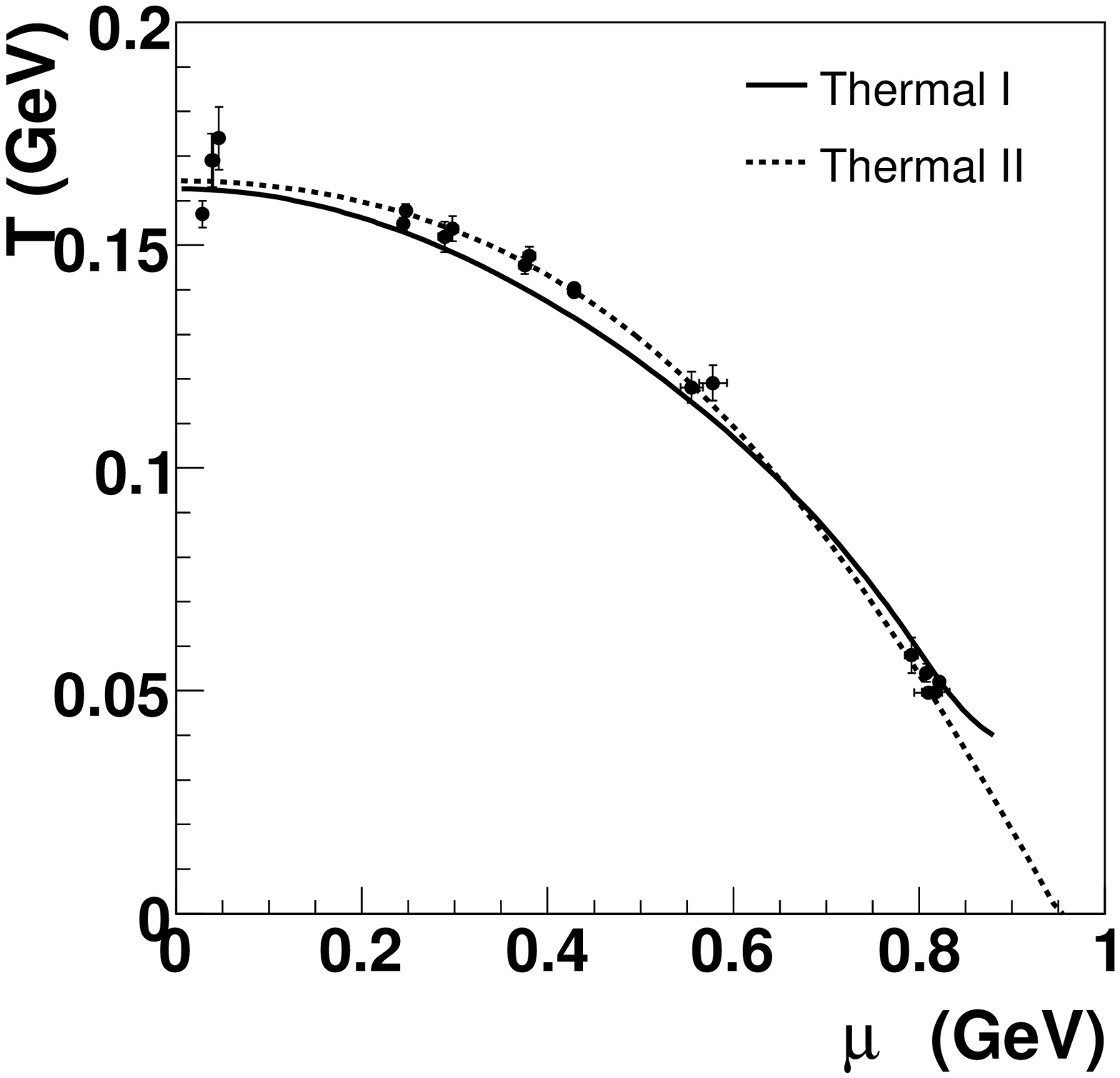}
\caption{
(left) Net baryon density for central A+A collisions at three CM energies,
from Ref.\cite{Bearden:2003hx}.  
(right) Compilation of extracted values of $T$ and $\mu_B$, as presented
in Ref.~\cite{Cleymans:2005km}.  Two parametrizations of $T(\mu_B)$ are
shown, as described in the text.
\label{fig:paperEnergy}
}
\end{center}
\end{figure}

It was discussed above that the leading particles in $p+p$ collisions  
(generally thought to be the initial protons) 
take a fraction of the initial energy ($x_F = 2p_z/\sqrt{s}$) with
a distribution that is flat over most of $x_F$ (leading to $\langle x_F \rangle
\sim 1/2$), suggesting that $dN/dy \sim exp(y)$.
Contrary to this, measurements of the net-baryon ($p-\overline{p}$) $dN/dy$
in A+A collisions over a wide range of $\sqrt{s_{NN}}$ find that
the net baryon density ``piles up'' at mid-rapidity at low energies
and probably peaks at around $y\sim 4$ at RHIC energies~\cite{Bearden:2003hx}.  
The interplay between the net baryon density (which is conserved) 
and the rest of the particle production
which is produced mainly by the freezeout of the strongly-interacting
matter, requires a baryochemical potential in thermal fits to the
yields of various hadron species~\cite{Cleymans:1998fq}.  
This is shown in the right
panel of Fig.~\ref{fig:paperEnergy} (from Ref.~\cite{Cleymans:2005km}) 
and is found to increase with
decreasing beam energy, as the overall net baryon density increases.

The depletion of the net-baryon density near mid-rapidity
has been called ``transparency'' by several authors.
This interpretation persists despite no corroborative 
evidence that the final distribution 
is simply from slowing down the initial-state baryons.  
Conversely, if one looks at the entropy alone, e.g. comparing
A+A to $e^+ e^-$ as was done in the previous section, one could 
arrive at the conclusion that the energy was fully stopped.
It is possible that the initial baryons are carried along by
the strong longitudinal expansion implied by Landau's initial conditions
(which is, ironically, much weaker than that described by Bjorken's
hydrodynamical model~\cite{Bjorken:1982qr}).  
Thus, at the very least, one should consider
the net baryon $dN/dy$ as the ``net'' rapidity distribution of the
initial state baryon excess, after both dynamical stopping 
and re-acceleration stages.

\begin{figure}[t]
\begin{center}
\includegraphics[width=140mm]{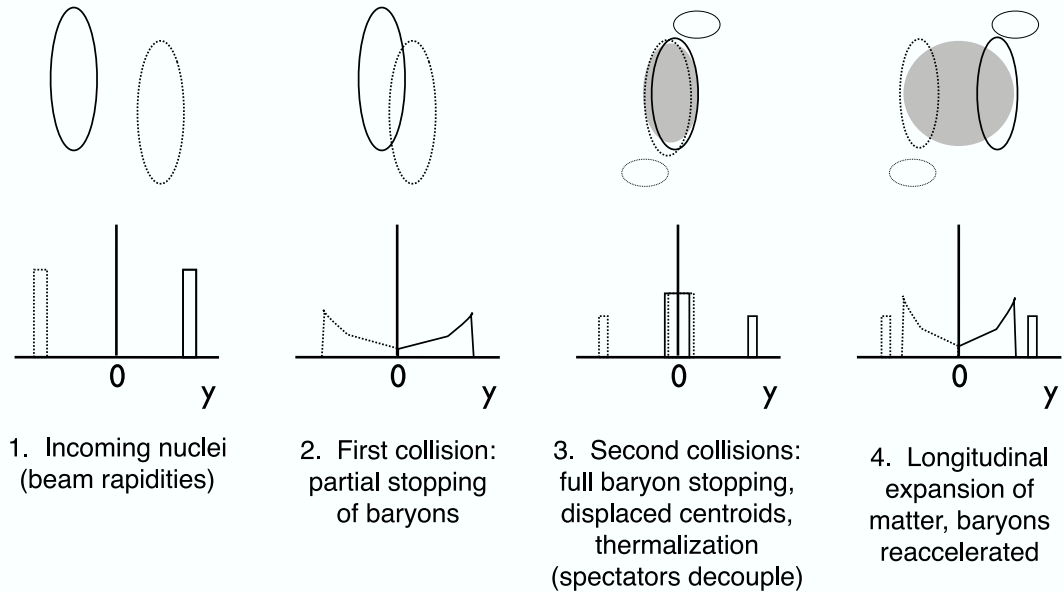}
\caption{
The ``Fireball Sandwich'' scenario for two colliding nuclei. 
\label{fig:stopping-sequence}
}
\end{center}
\end{figure}

As an example of a physical scenario which could generate the
BRAHMS results, consider the formation of a ``Fireball Sandwich''.
It is based on the simple assumption that a single collision of
a baryon on a target attenuates half of it's energy (as seen in
the comparison of the total entropy produced in $p+p$ 
reactions compared with $e^{+} e^{-}$) while practically all of the energy
is dissipated in the next collision -- a statement suggested, but not
completely proved, by the flat centrality dependence of $N_{ch}/(2N_{part})$
in A+A collisions which is 40\% higher than $p+p$~\cite{Back:2006yw}.  
If this is the case, then while the first collision of
each incoming baryons in each nucleus is not sufficient to stop it,
the second (or perhaps third) might be. 
This will lead to both nuclei being fully stopped in
the initial state of the collision (as needed for Landau's hydrodynamic
model to work) but each cluster of nucleons will be displaced from
$z=0$ along the original direction of motion. It is not difficult to
assume that the deposited energy will be mainly at $z=0$ and thus
behind the two walls of baryons on the outside of the reaction zone.
This creates a ``sandwich'' configuration, with layers of the
incoming net baryons tending to be near the edges of the hot 
thermalized matter.  These steps are illustrated (crudely) in 
Fig.~\ref{fig:stopping-sequence}.

A key feature of hydrodynamic models is the strong correlation between
the final state rapidity and the initial position of a fluid element
relative to the light cone~\cite{Belenkij:cd}.  
Fluid elements closer to the
outside edge of the initial thermalized region will be pushed from
behind and hadronize at large rapidities.  Conversely, fluid elements
near the center of the reaction zone will (unsurprisingly) hadronize
near mid-rapidity.  Thus, the Fireball Sandwich scenario naturally
explains how a net-baryon $dN/dy$ resembling ``transparency''
could be generated from completely-stopped baryons.  Of course,
it does not explain the dynamical mechanisms behind baryon stopping, 
but it might
allow experimental measurements to gain insight into the space-time
structure of stopping in A+A collisions.

If the baryons are ultimately in the initial reaction, and the system
is locally strongly-coupled, then it is
almost inevitable that the will participate in the overall chemistry
of the reaction, including the total entropy -- and thus the
total multiplicity.  This can be seen by writing down the
Gibbs potential for a system with a conserved baryon number:
\begin{equation}
G=E+PV-TS = \mu_B N_B
\end{equation}
This rearranges to
\begin{equation}
S = \frac{E+PV}{T} - \frac{\mu_B N_B}{T} \equiv S_0 - S_B.
\end{equation}
This shows that the presence of a baryochemical potential 
{\it reduces} the total entropy, and thus the total multiplicity.
If one scales out the total participating baryon number $N_B = N_{part}$
then the change in $N_{ch}/(N_{part}/2)$ is
\begin{equation}
\Delta N_{ch} \propto \frac{S_B}{N_B} \propto \frac{\mu_B}{T}
\end{equation}

\begin{figure}[t]
\begin{center}
\includegraphics[width=60mm]{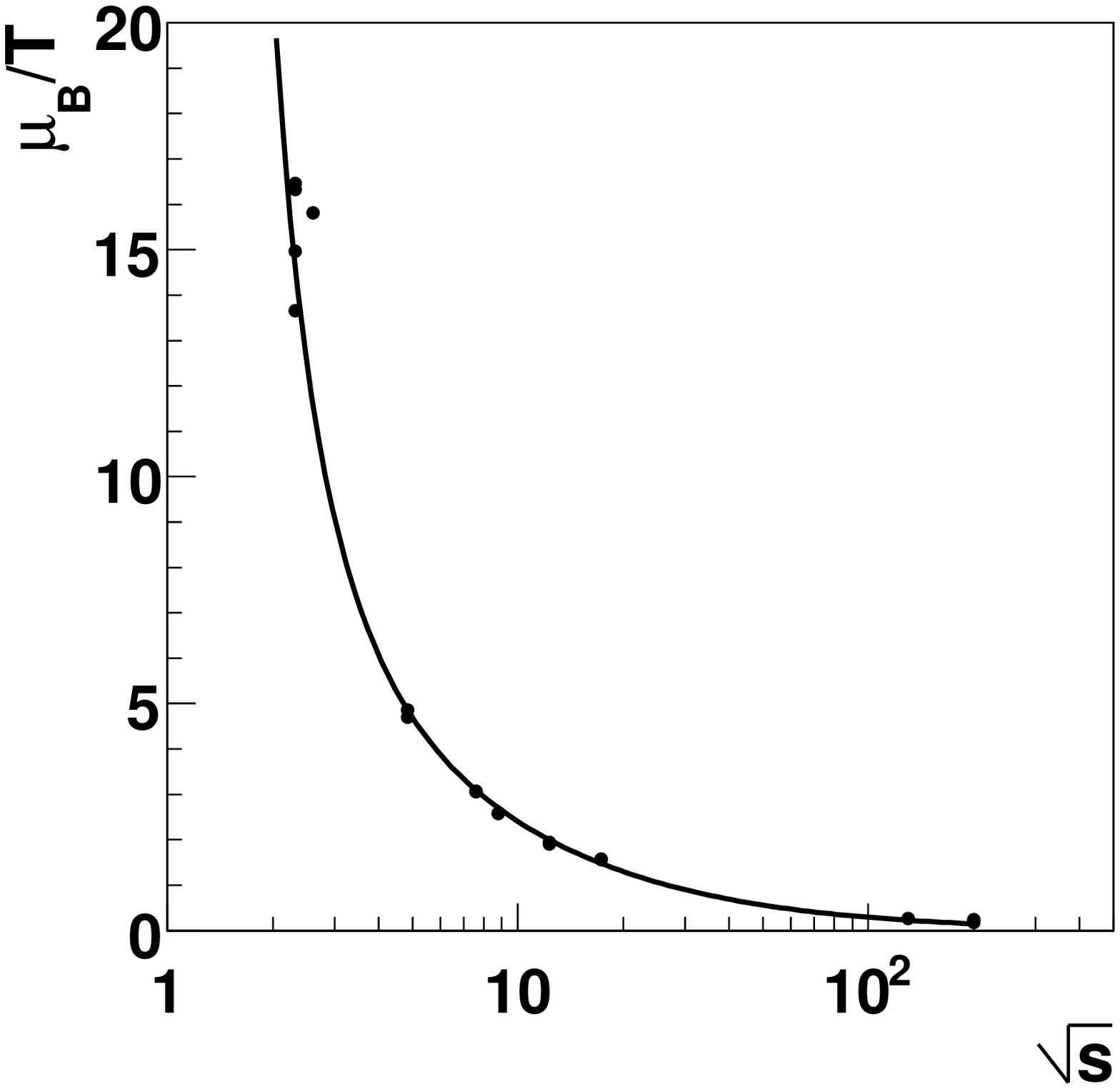}
\includegraphics[width=90mm]{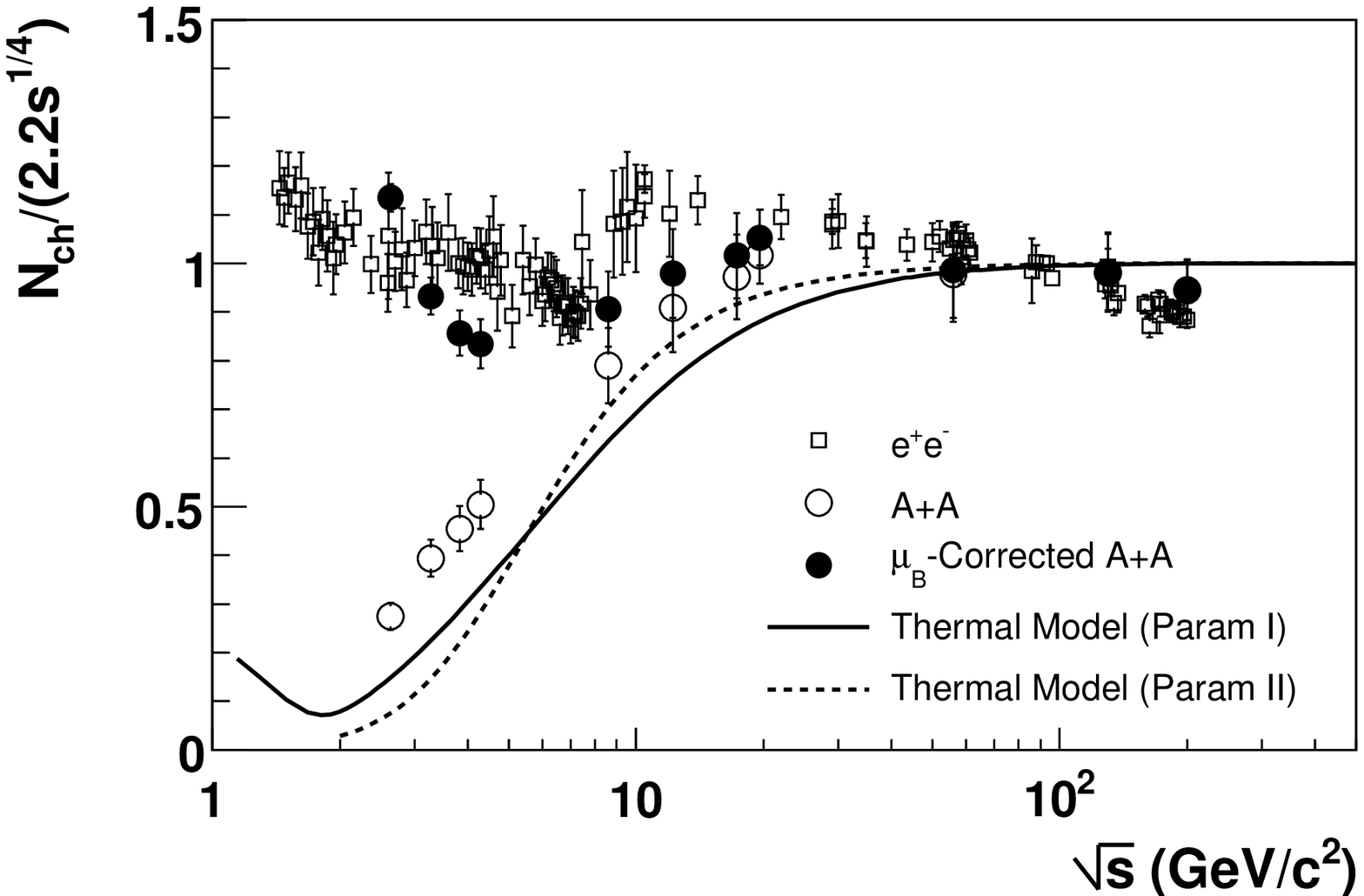}
\caption{
(left) $\mu_B/T$ vs. $\sqrt{s}$ extracted from data, as compiled in
Ref.\cite{Cleymans:2005km}. (right) Comparison of $e^+ e^-$ and
A+A data with two methods of estimating the effect of the net baryon
density, 1) ``correcting'' the A+A data to approximate $\mu_B=0$ and
2) direct thermal model calculations using $\mu_B(\sqrt{s})$ and
$T(\mu_B(\sqrt{s}))$ with two parametrizations.
\label{fig:ee_AA_mub_thermal}
}
\end{center}
\end{figure}

Of course, applying this formula to real data requires a ``baryon
free'' reference system for $N_{ch}$.  It has been shown above
that $p+p$ and $e^{+} e^{-}$ show the same energy dependence
over a large range in $\sqrt{s}$, if one corrects for the 
leading particle effect.  However, the presence of the leading
baryons in the $p+p$ final state might make it a more complicated
reference system, suggesting the use of $e^{+} e^{-}$ alone.
But once this reference system is chosen, there are (at least) two 
strategies for understanding the role of baryon number on entropy 
production.  One is to ``correct'' the A+A data for the presence
of a non-zero $\mu_B$, with help from thermal model phenomenology.  
The other is to do a straightforward
thermal model calculation to see how the entropy density
is modified as a function of $T$ and $\mu_B$ and translate this
into a ``suppression'' of $N_{ch}$ as a function of $\sqrt{s}$.

To correct the experimental data, it is only needed to calculate
$\mu_B/T$ as a function of $\sqrt{s}$ and then find the constant 
of proportionality (assumed to be energy independent) to convert
this to an inclusive charged-particle multiplicity.  
If $S/N = 4$ and $N/N_{ch} = 3/2$ (which is 
trivially true for a Boltzmann gas of massless pions), then
the conversion factor between $S/N$ and $N_{ch}/(N_{part}/2)$ is
exactly 6. Full thermal model calculations including strong decays
give a comparable factor of 7.2, only 15\% different than the
trivial estimate.  Thus, one adds $\Delta N_{ch} = (2./7.2) \mu_B/T$
to the measured value of $N_{ch}/(N_{part}/2)$
using the $\mu_B$ and $T$ extracted at each energy.  For simplicity,
we use a parametrization of $\mu_B(\sqrt{s}) = 1.2735/(1+0.2576\sqrt{s})$
extracted from the data shown in Fig.~\ref{fig:paperEnergy}.  
To convert $\mu_B$ into $T$, we use two parametrizations of $T(\mu_B)$,
one assuming that $\langle E \rangle / \langle N \rangle = 1 GeV$
(``Thermal I'', adapted from Ref.~\cite{Cleymans:1998fq}) 
and one using a third-order polynomial in $\mu^2_B$ (``Thermal II'').  

The direct calculation method is based on the formula:
\begin{equation}
\frac{2}{N_{part}} \frac{N^{A+A}_{ch}}{N^{e^+e^-}_{ch}} = 
\frac{C^{A+A}}{C^{e^+e^-}}
\frac{V^{A+A}}{V^{e^+e^-}}
\frac{s(T,\mu_B)}{s_0}
\end{equation}
where the formulae above are used to convert $\sqrt{s}$ to $T$
and $\mu_B$, and it is assumed that $C^{A+A}=C^{e^+e^-}$
and $V^{A+A}=V^{e^+e^-}$.  While assuming an equivalent conversion
from entropy to multiplicity seems natural, equating the
hadronization volume in $e^+ e^-$ and the volume per participant
pair in A+A seems less so.  It may well be a corollary result of this
work that it is not necessary to assume otherwise.
In any case, this calculation predicts the 
entropy suppression as a function of $\sqrt{s}$ just based on
experimental fits to ratios of particle yields.  Thus, it has
no free parameters and provides an approach complementary
to the correction method.

The final results for both the correction method and the direct
thermal model calculation are shown in Fig.\ref{fig:ee_AA_mub_thermal}.
The difference between them is a reasonable
estimate for the systematic uncertainties on these phenomenological
approaches.  It is observed that the corrected A+A data more
or less falls in line with the $e^+e^-$ data, and that the
direct thermal model calculations (a purely theoretical
calculation) qualitatively describe the 
suppression of the total multiplicity in A+A relative to $e^+ e^-$
(a ratio based purely on experimental data).  While the agreement
is certainly not perfect, no attempts have been made to improve it
by introducing {\it ad hoc} correction factors.

\begin{figure}[t]
\begin{center}
\includegraphics[width=110mm]{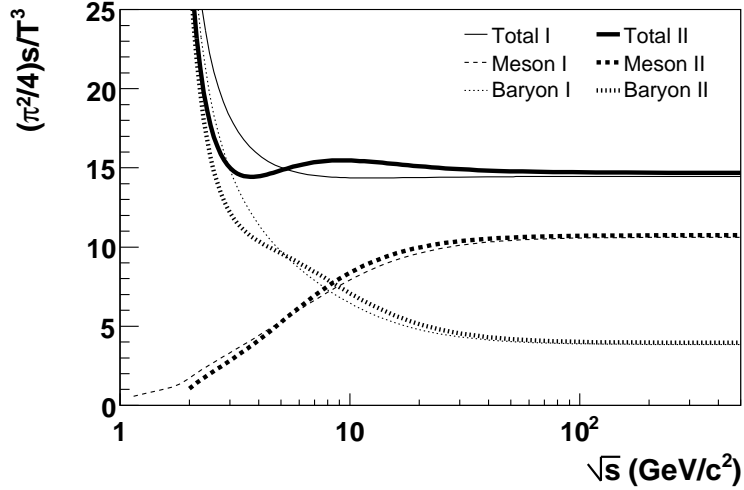}
\caption{\label{fig:DOF-band-3} The entropy density divided by $T^3$ as a function of $\sqrt{s}$ for mesons only, baryons only, and the total.  Again,two parametrizations are shown.  The scaling by $(4/\pi^2)$ transforms $s/T^3$ into the number of degrees of freedom of a massless relativistic Boltzmann gas. }
\end{center}
\end{figure}

It should be pointed out that all of the results shown in this
work are for inclusive charged hadrons.
This tends to be by necessity in the experiments represented
here, which only measure inclusive charged particles.  However,
there are theoretical reasons for including all particles, and not
choosing only mesons, as is often done in the literature.
Fig.~\ref{fig:DOF-band-3} shows a plot from Ref.~\cite{Cleymans:2006qe} 
performed for the
two parametrizations, but which is based on a figure shown in Ref.
It shows that the entropy per $T^3$, which is equal to the number
of degrees of freedom in a massless relativistic Boltzmann gas when multiplied
by $\pi^2/4$, is constant above $\sqrt{s}=3 GeV$ if one considers
contributions from both baryons and mesons.  It is manifestly
not constant with $\sqrt{s}$ if one chooses one or the others.
In particular, the meson sector rises very quickly and then plateaus,
something seen in the NA49 ``kink'' plot~\cite{Afanasiev:2002mx}.

\section{Summary and Outlook}

By showing how the low energy $A+A$ data can be made compatible
with $e^+ e^-$ data simply by considering the net baryon density,
this completes, in principle, the ``unification'' of the systematics 
of charged particle production in high energy multiparticle reactions.
From the comparisons of data shown above, one can glean three
``rules of thumb'' to understand the difference between the various
systems.
\begin{itemize}
\item The effective energy has a direct relationship to the entropy
(observed in $e^+ e^-$ vs. $p+p$ and then in comparison to A+A)
\item A stopped net baryon density suppresses the total entropy
(observed in $e^+ e^-$ vs. $A+A$ at low energies)
\item These connections are only seen when one uses $N_{ch}$,
i.e. one does not choose a particular measure of entropy.
\end{itemize}

It has been proposed here that the concepts and even the calculations
of Landau's hydrodynamic model can explain the relevance of these
rules.  Using rapid equilibration and hydrodynamic evolution as
guiding concepts to describe the ``bulk'' physics in A+A as well
as $e^+ e^-$ and $p+p$ seems to have qualitative and quantitative value,
and some new predictions for LHC energies have been given here.
Arguing that $e^+ e^-$ already has a comprehensive theory
may lead to missing insights to be gained from asking why
apparently different theories give similar results (e.g. 
why some pQCD calculations agree parametrically with Landau
hydro).

Whatever the theoretical situation,
detailed measurements of similar observables at higher energies or
at high $\mu_B$ should provide crucial new information.
The LHC will provide $p+p$ and A+A simultaneously, 
and FAIR at GSI will be specifically devoted to systems with
large net baryon stopping and thus high $\mu_B$.
In the high $\sqrt{s}$
sector, one will be testing the abilities of the system to thermalize,
or not, on astoundingly short timescales of $O(10^-3 fm/c)$.  In the high
$\mu_B$ sector one may be able to explore the systematics
of baryon stopping to understand the mechanisms of energy deposition.
Ultimately, one would like to understand all of this
physics in relation to the microscopic processes suggested by QCD.
In the meantime, the elegant structure of the data itself may well point
theory in completely new directions or suggest unexpected
connections between various techniques.

\end{document}